\documentclass[10pt]{article}
\usepackage{latexsym}
\usepackage{amssymb}
\usepackage{amsmath}
\usepackage[dvips]{graphicx}

\title{Exact solutions to $D=2$ \\ Supersymmetric Yang-Mills Quantum Mechanics \\ with $SU(3)$ gauge
group\footnote{Based on the talk given at the XLIX Cracow School of Theoretical Physics on 8 June 2009}}
\author{Piotr Korcyl\thanks{e-mail address: korcyl@th.if.uj.edu.pl} \\ \small{\emph{M. Smoluchowski Institute of Physics, Jagiellonian
University}} \\ \small{\emph{Reymonta 4, 30-059 Krak\'{o}w,
Poland}}}
\date{\today}

\begin{document}

\maketitle

\begin{abstract}
In this article we present the cut Fock space approach to the $D=d+1=2$, Supersymmetric Yang-Mills Quantum Mechanics (SYMQM).
We start by briefly introducing the main features of the framework. We concentrate on those properties of the method which
make it a convenient set up not only for numerical calculations but
also for analytic computations. In the main part of the article a sample of results are discussed,
namely, analytic and numerical analysis of the $D=2$, SYMQM systems with $SU(2)$ and $SU(3)$ gauge symmetry.
\end{abstract}

\section{Introduction}
\label{sec. Introduction}

Supersymmetric Yang-Mills Quantum Mechanics (SYMQM) turned out to be not only a class of supersymmetric systems
possessing some interesting physical features, but also to play an important r\^ole in many areas of theoretical physics.
Among the problems where SYMQM are relevant, the two most notable examples are: their relation to a particular limit of M-theory \cite{mtheory},
and the description of regularized dynamics of relativistic quantum membranes and supermembranes \cite{hoppe}\cite{dewit}.
Hence, an efficient way of
investigating the spectra of SYMQM with various gauge groups and defined in spaces with different dimensionality would be of importance.
%We would like to convince the Reader that the cut Fock space approach may be such a method.

In short, $D=2$, SYMQM are supersymmetric, $D=2$ dimensional, Yang-Mills quantum field theories reduced to one point in space.
Therefore, the original local gauge symmetry is transformed into a global symmetry of the quantum mechanical system.
The difficulty of solving these systems even in the simplest cases comes from the singlet constraint which is the remnant of the Gauss law.

%The physical Hilbert space with the singlet constraint imposed is composed exclusively
%of states invariant under the global symmetry.
The cut Fock space approach \cite{wosiek1} was proposed few years ago as a nonperturbative way of investigating
numerically the SYMQM systems.
%It incorporates the singlet constraint by constructing the Fock basis out of
%the most general set of linearly independent, gauge invariant states.
The energy eigenstates are constructed as linear combinations
of physical basis states and thus are gauge invariant by construction.
The approach allowed the calculation of low lying eigenenergies and their eigenstates through
a numerical diagonalization of the Hamiltonian matrix.
Its main application was the study of $D=2$, SYMQM \cite{wosiek2}\cite{wosiek3}\cite{doktorat_macka}, but it was also generalized
to the $D=4$ case \cite{wosiek5}. In principle
the method can be used in numerical investigations of systems with any gauge group and in any dimension.
Nevertheless, we would like to stress in this article that it is also convenient
for analytic treatment.
The cut Fock space approach provides a systematic way of controlling the Fock basis in terms of which the energy eigenstates can be decomposed.
As an example, we will present such decomposition of the solutions for two simple systems.

The paper is composed as follows. We start by briefly describing the cut Fock space approach, concentrating on the construction of the Fock basis.
Then, we present the numerical algorithm together with the numerical spectra of the $D=2$, SYMQM with $SU(3)$
gauge group in the sectors with $n_F=0$ and $n_F=2$ fermionic quanta.
In the third part of the paper, we derive analytically the bosonic solutions of the $D=2$, SYMQM Hamiltonians with $SU(2)$ and $SU(3)$ gauge groups.
We conclude by indicating possible directions of further studies.

\section{The framework}
\label{sec. Cut Fock basis framework}

For the reasons of simplicity the framework will be presented in the context
of $D=2$, supersymmetric Yang-Mills quantum mechanics with the $SU(N)$
gauge group. Nevertheless, the method is more flexible and systems with other gauge groups
as well as in higher dimensional spaces can be studied.
Particularly, creation and annihilation
operators introduced in the following
subsection can be labeled by additional spatial indices and
transform in some representation of the $SO(d)$ group.

%In the following we introduce the basic degrees of freedom, and the variables which will describe them. Then we present the construction
%of the Fock basis. Finally, we briefly describe the method for calculating the eigenenergies and eigenvectors.

\subsection{Basic degrees of freedom}

%The Hamiltonian investigated in this article is the simplest one corresponding to a set of free particles on a line\footnote{We adopt the notation when a repeated index is assumed to be summed over.}
%\begin{equation}
%H = \frac{1}{2} p_a p_a = \textrm{tr} \ p^2
%.
%\label{eq. The Hamiltonian}
%\end{equation}
A two-dimensional SYMQM system is described \cite{claudson} by a bosonic variable $\phi_A$
and a complex fermion $\lambda_A$, where $A$ is a color index.
Being remnants of the gauge field in the original field theory, the bosonic and fermionic variables
transform in the adjoint representation of the $SU(N)$ group.
Thus, the system contains $N^2-1$ bosonic %degrees of freedom
%, described by $\phi_A$ and their conjugate momenta $\pi_A$,
and $N^2-1$ fermionic degrees of freedom.
%, described by creation and annihilation operators, $f_A^{\dagger}$ and $f_A$.

The characteristic feature of SYMQM is that its Hilbert space is composed of states invariant under the $SU(N)$ group. We would
like to incorporate this constraint in the approach from the beginning.
Hence, we introduce a matrix notation, in which any singlet can be written in terms
of traces of appropriate matrices \cite{maciek1}.
We define
\begin{eqnarray}
\phi_{i,j} = \sum_{A=1}^{N^2-1} \phi^{}_A T^A_{i,j}, \quad && \quad \pi_{i,j} = \sum_{A=1}^{N^2-1} \pi^{}_A T^A_{i,j}, \nonumber \\
f^{\dagger}_{i,j} = \sum_{A=1}^{N^2-1} f^{\dagger}_A T^A_{i,j}, \quad && \quad f_{i,j} = \sum_{A=1}^{N^2-1} f^{}_A T^A_{i,j}, \nonumber
\end{eqnarray}
where $T^A_{i,j}$ are the generators of the $SU(N)$ group in the fundamental representation, $i,j = 1, \dots, N$.
Now, all operators
become operator valued matrices. Particularly, the
gauge-invariant occupation number operators can be defined as
\begin{equation}
\textrm{tr} \ (a^{\dagger} a) = \sum_{A=1}^{N^2-1} a^{\dagger}_A a^{}_A, \qquad
\textrm{tr} \ (f^{\dagger} f) = \sum_{A=1}^{N^2-1} f^{\dagger}_A f^{}_A,
\label{eq. occupation number operators}
\end{equation}
where we have introduced standard commuting creation and annihilation operators $a^{\dagger}_A, a^{}_A$
defined by $\phi^{}_A = \frac{1}{\sqrt{2}}(a^{}_A+a_A^{\dagger})$ and  $\pi^{}_A = \frac{1}{i\sqrt{2}}(a^{}_A-a_A^{\dagger})$. The
sum over the adjoint indices ensures the invariance of these operators under the $SU(N)$ transformations.

In the following we will consider the $D=2$, SYMQM systems with $SU(2)$ and $SU(3)$ gauge groups. Their Hamiltonians read respectively \cite{claudson},
\begin{equation}
H = \frac{1}{2} \pi_A \pi_A = \textrm{tr} \ (a^{\dagger}a) + \frac{3}{4} -\frac{1}{2} \big( \textrm{tr} \ (a^{\dagger} a^{\dagger}) - \textrm{tr} \  (aa) \big),
\label{eq. hamiltonian su2}
\end{equation}
\begin{equation}
H = \frac{1}{2} \pi_A \pi_A = \textrm{tr} \ (a^{\dagger}a) + 2 -\frac{1}{2} \big( \textrm{tr} \ (a^{\dagger} a^{\dagger}) - \textrm{tr} \ (aa) \big).
\label{eq. hamiltonian su3}
\end{equation}

\subsection{Fock basis}

The fundamental part of the approach is a systematic and recursive construction of the Fock basis.
Fock states are eigenstates of some occupation number operators, and
in the case of SYMQM models, we choose them to be eigenstates of the
gauge-invariant occupation number operators eq.\eqref{eq. occupation number operators}.

It is convenient when the Fock basis states can be labeled by as many quantum numbers conserved by the Hamiltonian as possible.
Hence, %we consider only gauge invariant states as candidates for the basis states.
since for most of systems the fermionic occupation number is conserved, one usually constructs
the Fock basis independently in each subspace
of the Hilbert space with a definite fermionic occupation number.
As far as the bosonic occupation number is concerned, it is in general not conserved.
Nevertheless, we will further divide the fermionic sectors into subspaces with a given number of bosonic
quanta, in order to facilitate the recursive approach.

In the following we introduce the concept of bosonic elementary bricks, which are necessary to obtain the Fock basis in the bosonic
sector. Then, we will proceed in full analogy with the fermionic sectors.

\subsubsection{Bosonic elementary bricks}

We define the set of \emph{bosonic elementary bricks} as the set of $N$ linearly independent
single traces of bosonic creation operators. Traces with more than $N-1$ operators can be reduced by the Cayley-Hamilton theorem.
Table
\ref{tab. elementary_bricks} contains examples of such sets for $N=2$, $N=3$ and $N=4$.
\begin{table}[h!]
\begin{center}
\begin{tabular}{|c|c|c|}
\hline
$SU(2)$ & $SU(3)$ & $SU(4)$\\
\hline
$\textrm{tr} \ (a^{\dagger}a^{\dagger})$ & $\textrm{tr} \ (a^{\dagger}a^{\dagger})$ & $\textrm{tr} \ (a^{\dagger}a^{\dagger})$ \\
 & $\textrm{tr} \ (a^{\dagger}a^{\dagger}a^{\dagger})$ & $\textrm{tr} \ (a^{\dagger}a^{\dagger}a^{\dagger})$ \\
 &  & $\textrm{tr} \ (a^{\dagger}a^{\dagger}a^{\dagger}a^{\dagger})$ \\
\hline
\end{tabular}
\caption{Elementary bosonic bricks for $SU(2)$,$SU(3)$ and $SU(4)$. \label{tab. elementary_bricks}}
\end{center}
\end{table}

Let us consider the set of states\footnote{We
adopt here the notation in which $(X)$ designs $\textrm{tr} \ (X)$.
We will use this notation only when it is self-evident.}
\begin{equation}
%\Big\{ |s_{n_B,0}(k_2, \dots, k_N) \rangle \Big\}_{\sum_{j=2}^{N}j k_j = n_B} = \Big\{ ( a^{\dagger 2})^{k_2} ( a^{\dagger 3})^{k_3} \dots ( a^{\dagger N})^{k_N}|0\rangle \Big\}.
\Big\{ ( a^{\dagger 2})^{k_2} ( a^{\dagger 3})^{k_3} \dots ( a^{\dagger N})^{k_N}|0\rangle \Big\}_{\sum_{j=2}^{N}j k_j = n_B} \equiv |\big\{n_B\big\}\rangle
\label{eq. bosonic fock basis}
\end{equation}
composed of the products of powers of elementary bosonic bricks acting on the Fock vacuum. One can show \cite{doktorat_macka} that
it spans the subspace of the Hilbert space with $n_B$ bosonic quanta.
In eq.\eqref{eq. bosonic fock basis}, we introduced a generalized notation in which $|\big\{n_B\big\}\rangle$ is a vector of all states with $n_B$ quanta.

Suppose that we have constructed such basis up to sectors containing less than $n_B$ bosonic quanta. The Fock basis in the sector with
$n_B$ bosonic quanta can be build as the sum of all states obtained
by the action of appropriate bricks on the already generated Fock basis states. Using the generalized notation this can be written in a compact form as
\begin{equation}
|\big\{n_B\big\}\rangle = \sum_{k=2}^{N} (a^{\dagger k}) |\big\{n_B-k\big\}\rangle. \nonumber
\label{eq. recursive fock basis}
\end{equation}
%Note that in general such states
%will not form an orthonormal set of states.
Note that the same state may appear in several copies,
differing in the order of successive bricks used to build it.
%Those duplicates will be treated as distinct states.
The basis is obtained once this redundancy is removed and the remaining states orthonormalized.

\subsubsection{Fermionic bricks}

In order to obtain the Fock basis in the fermionic sectors we must define bricks which contain fermionic creation operators.
The set of \emph{elementary fermionic bricks} can be defined in full analogy to the set of elementary bosonic bricks. We, thus, consider
all single traces with $n_F$ fermionic creation operators, which cannot be further reduced by the Cayley-Hamilton
theorem.
Subsequently, such set of elementary fermionic bricks must be enlarged by operators,
which are products of fermionic elementary bricks with smaller number of fermionic quanta
and contain $n_F$ fermionic creation operators in total.
The inclusion of these composite operators ensures that all possible
invariant contractions of $n_F$ fermionic and some bosonic creation operators are taken into account.
The enlarged set is called the set of \emph{composite fermionic bricks}.
%of all linearly independent fermionic bricks will be referred to as the set of
Its elements will be denoted by $C^{\dagger}(n_B, n_F, \alpha)$, where the first argument describes the number of bosonic creation
operators contained in $C^{\dagger}$, and the last argument is an additional label needed in the cases where $n_B$ and $n_F$ are not enough to
distinguish different operators. Table \ref{tab. su3_elementary_bricks} presents the set of composite fermionic bricks for the $SU(3)$ gauge group.

\begin{table}
\begin{center}
\begin{tabular}{|c|c|c|c|}
\hline
%$F=0$ &
$F=1$ & $F=2$ & $F=3$ & $F=4$\\
\hline
%$(a^{\dagger} a^{\dagger})$ &
$(f^{\dagger} a^{\dagger})$& $(f^{\dagger}f^{\dagger} a^{\dagger})$ & $(f^{\dagger}f^{\dagger}f^{\dagger})$ & $(f^{\dagger}f^{\dagger}f^{\dagger}f^{\dagger}a^{\dagger})$\\
%$(a^{\dagger} a^{\dagger} a^{\dagger})$ &
$(f^{\dagger}a^{\dagger} a^{\dagger})$ & $(f^{\dagger}f^{\dagger} a^{\dagger}a^{\dagger})$& $(f^{\dagger}f^{\dagger} f^{\dagger}a^{\dagger})$& $(f^{\dagger}a^{\dagger})(f^{\dagger}f^{\dagger}f^{\dagger})$\\
%&
& $(f^{\dagger} a^{\dagger}a^{\dagger}f^{\dagger}a^{\dagger})$& $(f^{\dagger} f^{\dagger}f^{\dagger}a^{\dagger}a^{\dagger})$& $(f^{\dagger}f^{\dagger}f^{\dagger}f^{\dagger}a^{\dagger}a^{\dagger})$\\
%&
& $(f^{\dagger}a^{\dagger})(f^{\dagger} a^{\dagger}a^{\dagger})$& $(f^{\dagger}a^{\dagger})(f^{\dagger} f^{\dagger}a^{\dagger})$&$(f^{\dagger}a^{\dagger}a^{\dagger})(f^{\dagger}f^{\dagger}f^{\dagger})$\\
%&
&& $(f^{\dagger}a^{\dagger}f^{\dagger}f^{\dagger}a^{\dagger}a^{\dagger})$ &$(f^{\dagger}a^{\dagger})(a^{\dagger}f^{\dagger}f^{\dagger}f^{\dagger})$\\
%&
&& $(f^{\dagger}a^{\dagger})(f^{\dagger}f^{\dagger}a^{\dagger}a^{\dagger})$ & $(f^{\dagger}f^{\dagger}a^{\dagger})(f^{\dagger}f^{\dagger}a^{\dagger})$\\
%&
&& $(f^{\dagger}a^{\dagger}a^{\dagger})(f^{\dagger}f^{\dagger}a^{\dagger})$ & $(f^{\dagger}a^{\dagger}a^{\dagger})(f^{\dagger}f^{\dagger}f^{\dagger}a^{\dagger})$\\
%&
&& $(f^{\dagger}a^{\dagger}a^{\dagger})(f^{\dagger}f^{\dagger}a^{\dagger}a^{\dagger})$ & $(f^{\dagger}f^{\dagger}a^{\dagger})(f^{\dagger}f^{\dagger}a^{\dagger}a^{\dagger})$\\
%&
&&& $(f^{\dagger}a^{\dagger})(f^{\dagger}a^{\dagger}a^{\dagger})(f^{\dagger}f^{\dagger}a^{\dagger})$\\
%&
&&& $(f^{\dagger}f^{\dagger}a^{\dagger})(f^{\dagger}a^{\dagger}f^{\dagger}a^{\dagger}a^{\dagger})$\\
\hline
\end{tabular}
\end{center}
\caption{Fermionic $SU(3)$ bricks. \label{tab. su3_elementary_bricks}}
\end{table}
In analogy to the bosonic case, we can define the set of states,
\begin{align}
%\Big\{ |s_{n_B,n_F}(k_2, \dots, k_N; n,\alpha) \rangle &\Big\}_{\substack{\sum_{j=2}^{N}j k_j = n_B - n\\ \\ \alpha \ = \ 1, \dots, d(n_F)}} = \nonumber \\
%\Big\{ |s_{n_B,n_F}(k_2, \dots, k_N; n,\alpha) \rangle &\Big\}_{\sum_{j=2}^{N}j k_j = n_B - n} = \nonumber \\
%&= \Big\{ C^{\dagger}(n, n_F, \alpha) C^{\dagger}(2,0)^{k_2} C^{\dagger}(3,0)^{k_3} \dots C^{\dagger}(N,0)^{k_N} |0\rangle \Big\},
\Big\{ C^{\dagger}(n, n_F, \alpha) (a^{\dagger 2})^{k_2} (a^{\dagger 3})^{k_3} \dots (a^{\dagger N})^{k_N} |0\rangle \Big\}_{\sum_{j=2}^{N}j k_j +n = n_B} \equiv |\big\{n_B,n_F\big\}\rangle
\label{eq. fermionic fock basis}
\end{align}
which after orthonormalization will give the basis in the subspace of Hilbert space with $n_B$ and $n_F$ bosonic and fermionic quanta, respectively.

\subsubsection{Correctness of the Fock basis}

One can show that the sets of
states eqs. \eqref{eq. bosonic fock basis} and \eqref{eq. fermionic fock basis}
provide indeed a good basis of the Hilbert space.
On one hand, its completeness follows from the fact that the states constructed with powers of bosonic and fermionic
bricks represent the most general contractions of invariant tensors with creation operators.
On the other hand, after removing the trivially linear dependent states, the linear independence
of the remaining ones can be check explicitly by calculating the Gram's matrix.
Obviously, the rank of the Gram's matrix corresponds to the number
of linearly independent states in a sector with given $n_B$ and $n_F$.
Fortunately, there exists also an independent way of calculating this number \cite{maciek2}.
It exploits the orthogonality of the characters of the $SU(N)$ group, and can be used as a crosscheck
that the Fock basis obtained through the recursive construction spans correctly the physical Hilbert space of SYMQM.\\

%Such a systematic construction of the Fock basis have several advantages.
%On one hand, it is a convenient set up for numerical calculations which will be described in section \ref{sec. Numerical results}.
%On the other hand, it facilitates the analytic treatment and enables one to find exact solutions, which will be demonstrated
%in section \ref{sec. Analytic solutions}.

\subsection{Extraction of approximate eigenenergies\\ and eigenstates}

\begin{figure}[!h]
\begin{center}
\includegraphics[width=5cm, height=5cm]{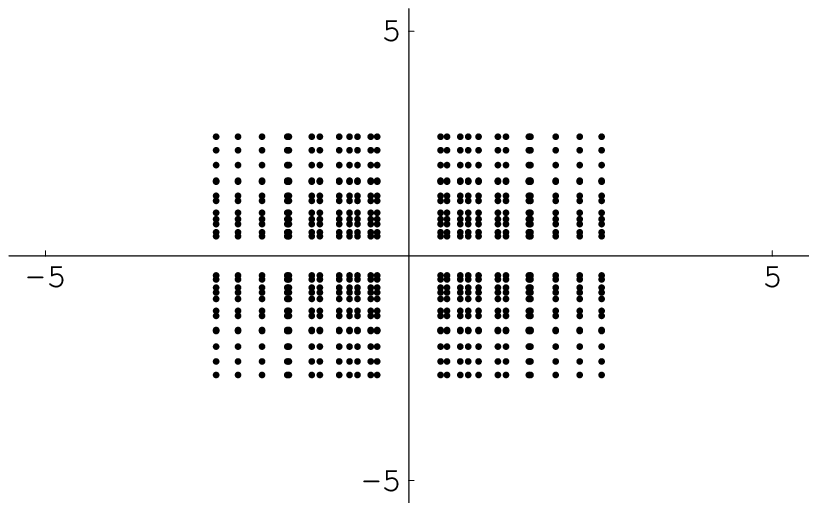}
\includegraphics[width=5cm, height=5cm]{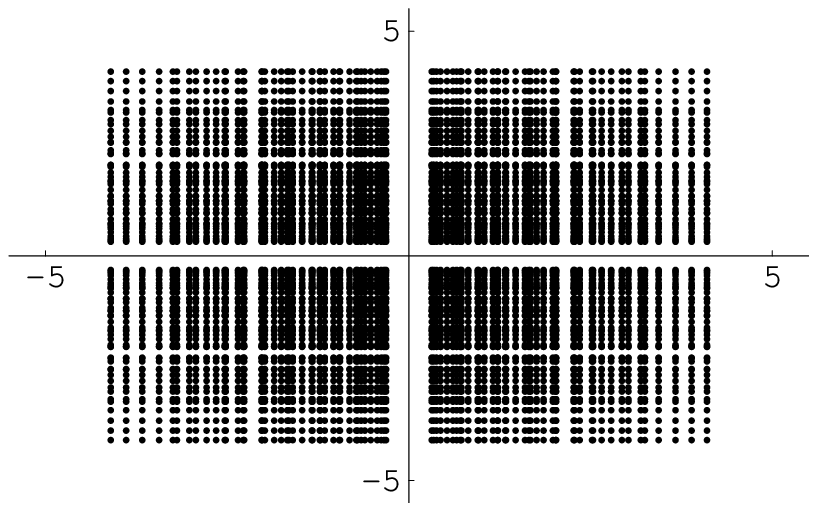}
\caption{Physical interpretation of the introduced cut-off. At finite cut-off the position and momentum operators have discrete spectra. As an example
eigenvalues of the pair of operators $\phi_1$ and $\phi_2$
as well as $\pi_1$ and $\pi_2$ are shown for different cut-offs
($N_{cut}=5,11$ for the left and right figures). With increasing cut-off
the points become more and more dense and tend to cover a bigger area on the plane. In the limit
of infinite cut-off the operators have a continuum spectrum and the whole plane is covered. \label{fig. cutoff}}
\end{center}
\end{figure}

Once the Hamiltonian operator is expressed
as an operator function of creation and annihilation operators,
its action is straightforward in the Fock basis.
However, the numerical analysis requires an additional step,
namely the introduction of a cut-off $N_{cut}$ (see figure \ref{fig. cutoff})
on the countably infinite Fock basis.
There are many ways to introduce such a
cut-off depending on the symmetries of the system.
A practical cut-off is a limit
on the total number of quanta contained in the Fock basis states.
Once the cut Hamiltonian matrix is obtained, its eigenvalues correspond to an approximation of the eigenenergies of the quantum
system, and its eigenvectors to the eigenstates.
Finally, calculations with several increasing $N_{cut}$ have to be performed
and the physical results extracted from the limit of infinite cut-off.
The properties of such a procedure were analyzed in \cite{maciek+wosiek}\cite{maciek3}\cite{praca_magisterska}\cite{korcyl2}, where
a different behavior of the eigenenergies corresponding to localized states and those corresponding to nonlocalized
states was observed. Therefore, the method offers a tool to distinguish these two types of states, and we will indeed
exploit this possibility when discussing the numerical results in subsection \ref{subsec. su3 model results}.

\section{Numerical results}
\label{sec. Numerical results}
In this section we briefly describe a recursive algorithm which can used to efficiently evaluate the matrix elements of the Hamiltonian operator.
Then, we present the spectra of $D=2$, SYMQM system with $SU(3)$ gauge group in the sectors with $n_F=0$ and $n_F=2$.

\subsection{Recursive numerical approach}

The recursive algorithm is based on relations connecting the desired matrix element of an operator to
simpler matrix elements of some other operators, which have been already evaluated at the earlier stage of calculations.
We will not describe here the full algorithm in details (for a full presentation of the approach see \cite{korcyl0}). Instead, we will
concentrate exclusively on the bosonic sector and
give one example of such recursive relation.

%However, before doing so, notice that the Fock basis states $| \big\{ n_B \big\} \rangle$ constructed recursively by eq.\eqref{eq. recursive fock basis},
%are not orthonormal. Moreover, such set of states contains multiple copies of the same state which differ
%only in the order of bricks used to construct them.
%Thus, in order to orthonormalize $| \big\{ n_B \big\} \rangle$ and to get rid of the linearly dependent states we evaluate the matrix of scalar
%products $S(n_B)$. Then, by numerical diagonalization of $S(n_B)$ we can obtain a matrix $R(n_{B})$ satisfying
%\begin{equation}
%R^T(n_{B}) S(n_B) R(n_{B}) = I , \nonumber
%\end{equation}
%where $I$ is the unity matrix which dimension
%is equal to the size of the subspace of the Hilbert space with $n_B$ bosonic quanta.
%Note, that such approach can be in a straightforward way
%generalized to fermionic sectors.

Let us now assume that we want to evaluate the matrix elements of the normally-ordered operator $O(n^{O}_{B})$ between
states containing $n_B$ and $n'_B$ quanta, $\langle \big\{n'_{B}\big\} |O(n^{O}_{B})| \big\{n_{B}\big\} \rangle$.
The argument of $O$ means that the difference between the number of creation
operators and the number of annihilation operators is $n^{O}_{B}$.
Therefore, the matrix elements $\langle \big\{n'_{B}\big\} |O(n^{O}_{B})| \big\{n_{B}\big\} \rangle$
will be nonzero only when $n'_{B} = n^{O}_{B} + n_{B}$.
The strategy to evaluate a matrix element of $O$ is
to drag $O$ over the operators constituting the state $| \big\{n_{B}\big\} \rangle$ so that $O$ annihilates the Fock vacuum.
We thus have
\begin{align}
\langle \big\{n'_{B}\big\} &|O(n^{O}_{B})| \big\{n_{B}\big\} \rangle  =
\Big( \langle \big\{n'_{B}\big\}| \big[O(n^{O}_{B}), (a^{\dagger p})\big] | \big\{n_{B}-p\big\}\rangle \nonumber \\
&+ \langle \big\{n'_{B}\big\}| (a^{\dagger p}) O(n^{O}_{B}) | \big\{n_{B}-p\big\}\rangle \Big) \cdot R (n_{B}),
\label{eq. correct bosonic recursion relation}
\end{align}
where the matrix $R(n_{B})$ is obtained from the matrix of scalar products $S(n_B) = \langle \big\{ n_B\big\} | \big\{ n_B \big\} \rangle$,
and is used to remove redundant basis vectors and orthonormalize the remaining ones. Thus, $R(n_B)$ satisfies
\begin{equation}
R^T(n_{B}) S(n_B) R(n_{B}) = I , \nonumber
\end{equation}
where $I$ is the unity matrix which dimension
is equal to the size of the subspace of the Hilbert space with $n_B$ bosonic quanta.
Note that  we have expressed the
desired matrix element through eq.\eqref{eq. correct bosonic recursion relation} in terms of matrix elements of operators between states with lower
number of quanta, which should have been already evaluated during some previous calculations. 
Hence, applying
successively such relation one can evaluate $\langle \big\{n'_{B}\big\} |O(n^{O}_{B})| \big\{n_{B}\big\} \rangle$. One can similarly organize the
calculations in the fermionic sectors.

%In principle the above algorithm can be used to systems defined in space of any dimensionality
%and it is applicable to systems with bosonic and fermionic
%polynomial interactions as well as to systems with discrete or continuum spectrum.

\subsection{$SU(3)$ model}
\label{subsec. su3 model results}

\begin{figure}[t]
\begin{center}
\includegraphics[width=0.6\textwidth,angle=270]{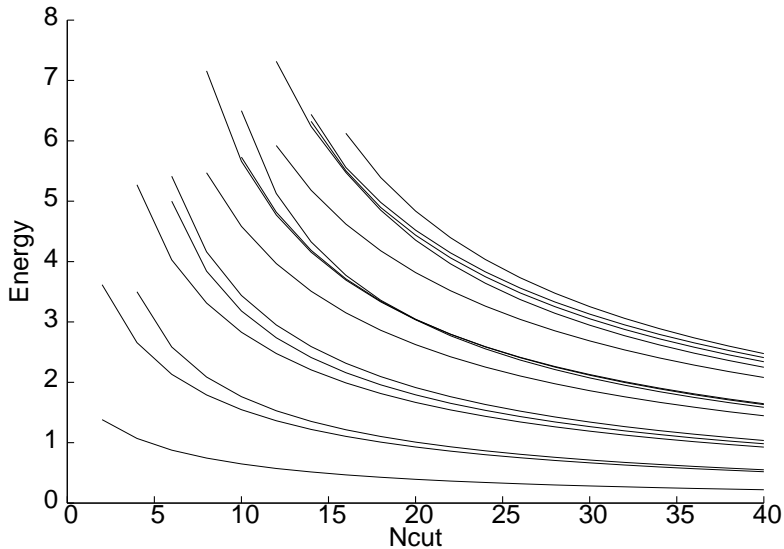}
\caption{Dependence of the eigenenergies on the cut-off for the $SU(3)$ model in the $n_F=0$ sector. \label{fig. spectrum1}}
\end{center}
\end{figure}

\begin{figure}[t]
\begin{center}
\includegraphics[width=0.6\textwidth,angle=270]{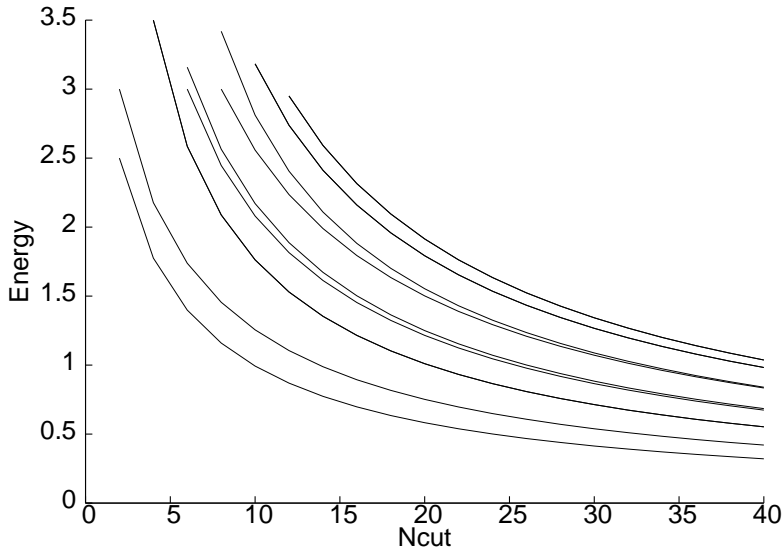}
\caption{Dependence of the eigenenergies on the cut-off for the $SU(3)$ model in the $n_F=2$ sector. \label{fig. spectrum2}}
\end{center}
\end{figure}

As an application of the above algorithm we present the spectra of the system given by the Hamiltonian eq.\eqref{eq. hamiltonian su3}.
Figures \ref{fig. spectrum1} and \ref{fig. spectrum2} contain the results in the sectors with $n_F=0$ and $n_F=2$, respectively.
We notice that in both cases all eigenenergies fall down to zero. Thus, we can conclude that
the corresponding eigenstates are nonlocalized.
It is a feature of the approach that the eigenvalues corresponding to such
states do not converge with the increasing cut-off.
Such behavior results
from approximating a plane wave by a finite set of localized harmonic oscillator eigenstates
and its dependence on $N_{cut}$ can be parameterized in a power like manner. A precise
procedure for obtaining the correct infinite cut-off limit of the energy
was described in \cite{maciek+wosiek}\cite{maciek3}\cite{korcyl2}. In particular, it was shown that the hyperbolic
fall off contains information about the dispersion relation. Note that for a finite $N_{cut}$ there
can be only a finite number of eigenenergies. With increasing $N_{cut}$ more and more new
eigenvalues should appear on plots figs. \ref{fig. spectrum1} and \ref{fig. spectrum2}, in the limit giving
the continuum spectrum of a free system. In the next section we will present analytic calculations which
enable one to completely reconstruct the above spectra, and thus provide a full understanding of their features.
%A full understanding of these figures will be given when the analytic solutions
%at finite cut-off will be described in the following section.

\section{Analytic solutions}
\label{sec. Analytic solutions}

Our procedure to derive exact solutions for the $D=2$, SYMQM models consists in three steps. We decompose a
general state $|E\rangle$ from the cut Hilbert space in the Fock basis. In this step a convenient parametrization of the Fock basis
is necessary. Then, we translate the requirement that $|E\rangle$ is an eigenstate of the appropriate Hamiltonian into
a recurrence relation on the decomposition coefficients. The last step is the solution of this recurrence relation
at finite cut-off and the investigation of the infinite cut-off limit of these solutions.

Before we deal with the Hamiltonian \eqref{eq. hamiltonian su3}, we will analyze a simpler model with the $SU(2)$ symmetry group,
\eqref{eq. hamiltonian su2}.
The solution of the $SU(2)$ model was first found by Claudson and Halpern \cite{claudson}.
It can be obtained from the generalized
solutions of Samuel \cite{samuel} and was also recently rederived by Trzetrzelewski \cite{doktorat_macka} by algebraic methods. This model is particularly simple because the
$SU(2)$ group is isomorphic with the $SO(3)$ group and can be parameterized by spherical coordinates.

\subsection{$SU(2)$ model}

\subsubsection{Recurrence relation}

At cut-off $N_{cut}$ the general state from the bosonic sector can be decomposed as
\begin{equation}
|E\rangle = \sum_{j=0}^{N_{cut}} a_j(E) \ \textrm{tr} \ (a^{\dagger 2})^j |0\rangle. \nonumber
\end{equation}
The eigenequation
\begin{equation}
H |E\rangle = E |E\rangle, \nonumber
\end{equation}
yields the recurrence relation for the $a_j(E)$ coefficients,
\begin{equation}
a_{j-1}(E) - \big( 2j + \frac{3}{2} - 2 E \big) a_{j}(E)  + (j+1) \big( j + \frac{3}{2}\big) a_{j+1}(E) = 0,
\label{eq.su2.recurrence.rel}
\end{equation}
which can be solve analytically as will be discussed in the following subsections.

\subsubsection{Finite cut-off solutions}

%At finite cut-off $N_{cut}$ o
One can show \cite{korcyl3} that
 eq.\eqref{eq.su2.recurrence.rel} admits $N_{cut}+1$ solutions.
The possible eigenenergies are given by the zeros of an appropriate associated Laguerre
polynomial\footnote{Laguerre
polynomials $\mathcal{L}_n^{\alpha}(x)$ are defined
as the solutions of the differential equation
\begin{equation}
x y'' + (\alpha +1 -x) y' + n y = 0, \nonumber
\end{equation}
and the orthogonality relation
\begin{equation}
\int_0^{\infty} \mathcal{L}_m^{\alpha}(x) \mathcal{L}_n^{\alpha}(x) x^{\alpha} e^{-x} dx = \delta_{m n}. \nonumber
\end{equation}
$L_n^{\alpha}(x)$ denotes the Sonine polynomials related to Laguerre polynomials by
\begin{equation}
L_n^{\alpha}(x) = \frac{1}{\Gamma(\alpha+n+1)} \mathcal{L}_n^{\alpha}(x). \nonumber
\end{equation}
},
\begin{equation}
L^{\frac{1}{2}}_{N_{cut}+1}(2E) = 0. \nonumber
\end{equation}
For each $E$ satisfying the above equation there exist an eigenstate with $N_{cut}+1$
decomposition coefficients are given by
\begin{equation}
a_j(E) = a_0 \Gamma(\frac{3}{2}) L^{\frac{1}{2}}_j(2E) \qquad 0 \le j \le N_{cut}, \nonumber
\end{equation}
where $a_0$ is some arbitrary constant.

\subsubsection{Infinite cut-off solutions}

In the infinite cut-off limit, $N_{cut} \rightarrow \infty$, the set of possible eigenenergies
is given by the whole real positive axis. This reflects the fact that
the physical spectrum is continuous.
Hence, for any real number $E$, there exists an eigenstate of $H$
which decomposition coefficients are given by
\begin{equation}
a_j(E) = a_0 \Gamma(\frac{3}{2}) L^{\frac{1}{2}}_j(2E) \qquad j \ge 0. \nonumber
\end{equation}
Therefore, the exact eigenstate can be written as,
\begin{equation}
|E\rangle = a_0 \Gamma(\frac{3}{2}) \sum_{j=0}^{\infty} L^{\frac{1}{2}}_j(2E) \ \textrm{tr} \ (a^{\dagger 2})^j |0\rangle. \nonumber
\end{equation}

\subsubsection{Reconstruction of wavefunctions}

In the case of the $SU(2)$ model one can explicitly check the correctness of the above solutions.
Let us denote by $\psi_n(r)$ the wavefuntion in the position representation
of the $n$-th basis state. $r$ is the radial variable which parameterizes
the $SU(2)$ group manifold.
Since we consider only $SU(2)$ invariant states, $\psi_n(r)$ do not depend on angular variables.
The $n$-th basis state being the eigenstates of the gauge-invariant particle number operator $\textrm{tr} \ (a^{\dagger} a)$, its wavefunction
satisfies the following equation
\begin{equation}
- \frac{1}{2}\Big( \frac{d^2}{dr^2} + \frac{2}{r} \frac{d}{dr} + 3 - r^2 \Big) \psi_n(r) = 2 n \ \psi_n(r). \nonumber
\end{equation}
Such equation can be solved yielding
\begin{equation}
\psi_n(r) = \alpha(n) e^{-\frac{r^2}{2}} (-1)^n 2^{2n+1} n! \mathcal{L}_n^{\frac{1}{2}}(r^2) + \beta e^{-\frac{r^2}{2}} \frac{1}{r} {}_1 F_1 (- \frac{2n+1}{2}, \frac{1}{2}, r^2), \nonumber
\end{equation}
where $\alpha(n)$ is some constant depending on $n$ and
${}_1 F_1 (a, b, z^2)$ is the Kummer's function of the first kind.
We are only interested in normalizable solutions, therefore
since the function $\frac{1}{z} {}_1 F_1(a,b,z^2)$ is singular at $z=0$,
we set $\beta=0$.
%Moreover,
%\begin{equation}
%\frac{1}{r} H_{2n+1}(r) = (-1)^n 2^{2n+1} n! \mathcal{L}_n^{\frac{1}{2}}(r^2).
%\end{equation}
%Thus
%\begin{equation}
%n(r) = \alpha  (-1)^n 2^{2n+1} n! e^{-\frac{r^2}{2}} \mathcal{L}_n^{\frac{1}{2}}(r^2).
%\end{equation}
Thus, $\psi_n(r)$ turns out to be
the wavefunction of a three dimensional harmonic oscillator carrying zero angular momentum.
In order to find explicitly the Claudson-Halpern solutions,
we write the eigensolution with energy $E$ as
\begin{align}
\langle r|E\rangle &= \sum_{j=0}^{\infty} \langle r| \textrm{tr} \ (a^2)^j \rangle \langle \textrm{tr} \ (a^{\dagger 2})^j | E\rangle = \nonumber \\
%& = \sum_{j=0}^{\infty} a_j(E) \phi(E) \langle r| \textrm{tr} \ (a^{\dagger 2})^n |0 \rangle
&= a_0 \Gamma(\frac{3}{2}) e^{-\frac{r^2}{2}} \phi(E) \sum_{j=0}^{\infty}
\frac{\alpha(j) 2^{2j+1} (-1)^j \ j!}{\Gamma(j+\frac{3}{2})} \mathcal{L}^{\frac{1}{2}}_j(2E) \mathcal{L}_j^{\frac{1}{2}}(r^2), \nonumber
\end{align}
where $\phi(E)$ is some function of the variable $E$ which was not determined by the recursion relation eq.\eqref{eq.su2.recurrence.rel}.
Choosing $\alpha(n) = 2^{-2j-1}$ and setting $2E=k^2$ we can use
a known formula for the sum of products of associated Laguerre polynomials of the same index \cite{abramowitz},
\begin{equation}
\sum_{j=0}^{\infty} \frac{(-1)^j j!}{\Gamma(j+\frac{3}{2})}\mathcal{L}^{\frac{1}{2}}_j(r^2) \mathcal{L}^{\frac{1}{2}}_j(k^2) = \frac{1}{2} \exp \big( \frac{k^2+r^2}{2} \big) \frac{1}{\sqrt{i k r}} I_{\frac{1}{2}} \big( \sqrt{i k r} \big). \nonumber
\end{equation}
Next, by exploiting some properties of the Bessel functions \cite{abramowitz}, and setting $\phi(E) = e^{-E}$,
we can transform the above result into
\begin{equation}
\langle r|E\rangle = \frac{a_0}{2} \Gamma(\frac{3}{2}) \sqrt{\frac{2}{\pi}} \frac{\sin (k r)}{k r}, \nonumber
\end{equation}
which is, up to a multiplicative factor, the Claudson-Halpern solution of the $SU(2)$ model.

\subsection{$SU(3)$ model}

%In this section we repeat the analysis described above for the model with $SU(3)$ gauge group.
%The results will be compared with the numerical results
%presented in section \ref{subsec. su3 model results}.

\subsubsection{Recursion relation}

We now decompose $|E\rangle$ in the Fock basis of the $SU(3)$ model
\begin{equation}
|E\rangle = \sum_{2j+3k \le N_{cut}} a_{j,k}(E) \ \textrm{tr} \ (a^{\dagger 2})^j \ \textrm{tr} \ (a^{\dagger 3})^k |0\rangle. \nonumber
\label{eq.expansion}
\end{equation}
%Let us denote the Hamiltonian eigenstate to the energy eigenvalue $E$ by $|E\rangle$,
%\begin{equation}
%H|E\rangle = E |E\rangle. \nonumber
%\label{eq.eigen equation}
%\end{equation}
For $|E\rangle$ being an eigenstate, $a_{j,k}$ must obey the following
recursion relation \cite{korcyl3},
\begin{align}
a_{j-1,k} - \big( 2j + 3k + 4 -2E \big) a_{j,k} &+ (j+1)(j+3k+4) a_{j+1,k} \nonumber \\
&+ \frac{3}{8}(k+1)(k+2) a_{j-2,k+2} = 0.
\label{eq.su3.recurrence.rel}
\end{align}
Notice that the first three terms are diagonal in the $k$ index and are similar to the recurrence
relation eq.\eqref{eq.su2.recurrence.rel}. The last term in eq.\eqref{eq.su3.recurrence.rel} mixes the
coefficients with different values of the $k$ index.
However, coefficients $a_{j,k}$ with even and odd $k$ remain not related.
Therefore, we can solve separately for the amplitudes $a_{j,2k}$ and $a_{j, 2k+1}$.

%The introduction of the cut-off $N_{cut}$ translated in terms of the coefficients $a_{j,k}$,
%constrains the set of variables to only those $a_{j,k}$ for which $2j+3k \le N_{cut}$.
%Moreover, we also impose that $\forall_{j<0, k<0} \quad a_{j,k} = 0$.

\subsubsection{Finite cut-off solutions}

For reasons of clarity we consider here only the situation when
the cut-off $N_{cut}$ is even and the solutions contain an even number,
$2m$, of cubic bricks $(a^{\dagger 3})$. The derivation of solutions with an odd number of cubic
bricks is similar.

It can be shown \cite{korcyl3} that the solutions
to eq.\eqref{eq.su3.recurrence.rel} can be classified
into several separate sets.
A solution belongs to the set $f_m$ if $a_{j,k} \equiv 0, k > 2m$
and $a_{j,k} \ne 0, k \le 2m$. In words, this means that the eigenstate
can be decomposed into basis states containing at most $2m$ cubic bricks.
$f_0$ is the simplest set of solutions, for which only $a_{j,0}$ are nonzero, i.e.
is only composed of bilinear bricks. Each set of solutions has its separate quantization
condition for the possible values of the $E$ parameter.

\begin{figure}[!h]
\begin{center}
\includegraphics[width=0.6\textwidth,angle=270]{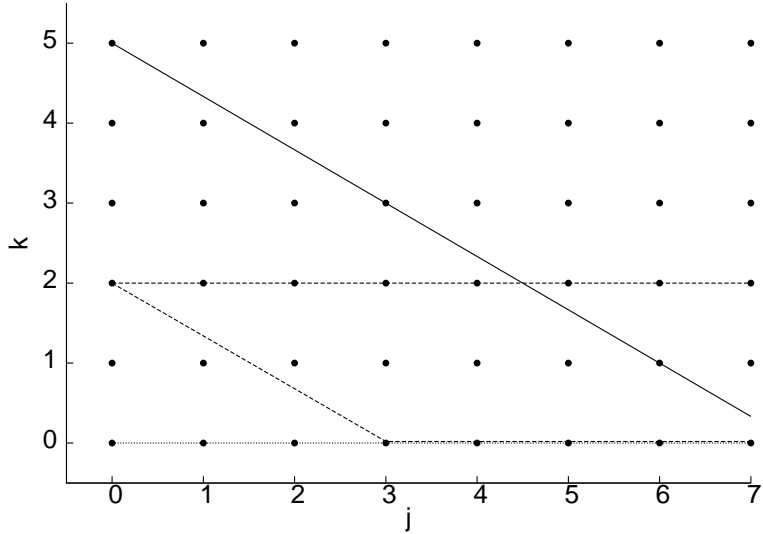}
\caption{The structure of the solutions of the recursion relation eq. \eqref{eq.su3.recurrence.rel}. Each dot represents
a coefficient $a_{j,k}$ with appropriate values of the $j$ and $k$ indices. The oblique, straight line denotes a cut-off with
a fixed number of quanta, here $N_{cut}=15$. The coefficients lying below and on this line are included
in the Fock basis, whereas coefficients lying outside are not. By increasing the cut-off we push this line to the right and include
more states into the cut Fock basis. The remaining lines represent the sets of amplitudes of
particular solutions of the recursion relation. The lowest, dotted line corresponds to the solution involving only
quadratic bricks i.e. the solution from the set $f_0$. The dashed triple represents a solution from the set $f_1$.
The two horizontal parts of the triple denote the amplitudes $a_{j,2}$ and $a_{j,0}$. The mixing of these amplitudes starts
at the number of quanta equal to 6, i.e. both the amplitudes $a_{0,2}$ and $a_{3,0}$ contain 6 quanta.
\label{fig. siatka}}
\end{center}
\end{figure}

We will now give the general form of solutions belonging to the set $f_m$.
The set $f_m$ contains $d_m \equiv \frac{1}{2} \big( N_{cut} - 6m) \big) + 1$
solutions with $E$ such that $L_{d_m}^{6m+3}(2E) = 0$. They can be written
as\footnote{We adopt a simplified notation in which $ \Big(\textrm{tr} \ (a^{\dagger 2}) \Big)^j \ \Big(\textrm{tr} \ (a^{\dagger 3}) \Big)^k |0\rangle \equiv |j,k\rangle$.} \cite{korcyl3}
\begin{align}
|E\rangle &= \sum_{j=0}^{\#f_m} L_j^{6m+3}(2E) \Big( |j,2m\rangle + \sum_{p=1}^{m} \Gamma^E(m,p) | j+3p,2m-2p\rangle \Big) \nonumber
\end{align}
with
\begin{equation}
\Gamma^E(m,p) = \prod_{t=p}^{m} - \frac{1}{3} \frac{t(2t-1)}{(2m+1)^2 - (2t-1)^2} \nonumber %= \frac{24^{p-m} \Gamma(-2p)}{\Gamma(-m-p) \Gamma(1+m-p)}. \nonumber
\end{equation}
In order to illustrate these formulas, let us present explicitly the solutions from the sets $f_0$ and $f_1$. They read, respectively,
\begin{align}
|E\rangle &= a_0 \sum_{j=0}^{d_0} L_j^3(2E) |j,0 \rangle, \nonumber \\
|E\rangle &= a_0 \sum_{j=0}^{d_1} L_j^9(E) \Big( |j,2\rangle - \frac{1}{24} |j+3,0\rangle \Big). \nonumber
\end{align}
The new feature of the $f_{>0}$ solutions is the degeneracy, which appears because
several states can contain the same total number of quanta, i.e. the equation $2j+3k=n_B$ can have several solutions.
Particularly, the degeneracy of the states containing 6 quanta, namely $|2,0\rangle$ and $|0,3\rangle$ is responsible
for the structure of the solutions from the set $f_1$.
Figure \ref{fig. siatka} demonstrates graphically the structure of these solutions.

%$m_{2p+1}+1$ solutions with $E$ such that $L_{m_{2p}+1}^{3(2p+2)}(4E) = 0$ are given by
%\begin{align}
%|E\rangle = \sum_{n=0}^{m_{2m+1}} L_n^{3(2m+2)}(4E) \Big( |n,2m+1\rangle &+ \Gamma^O(m,m-1) |n+3,2(m-1)+1\rangle + \nonumber \\
%&+  \dots + \Gamma^O(m,1) | n+3m,1\rangle \Big)
%|E\rangle &= \sum_{n=0}^{\#f_m} L_n^{3(2m+2)}(E) \Big( |n,2m+1\rangle +\sum_{p=1}^{m-1} \Gamma^O(m,p) | n+3p,2m-2p+1\rangle \Big) \nonumber
%\end{align}
%with
%\begin{equation}
%\Gamma^O(m,p) = 2^{1-2p}3^{p-m} \sqrt{\pi}
%\frac{\Gamma(m+\frac{1}{2})(m+p-1)!}{(m-1)!(m-p)! \Big( \Gamma(p+\frac{1}{2})\Big)^2}. \nonumber
%\end{equation}

The complete solution to the eigenvalue problem, i.e. the set of all eigenstates $\big\{ | E \rangle \big\}$ is given by the union
\begin{equation}
\big\{ | E \rangle \big\} = \bigcup_{m=0}^{\big\lfloor \frac{1}{6} N_{cut} \big\rfloor} f_m \ \cup \ \bigcup_{m=0}^{\big\lfloor \frac{1}{6} \big( N_{cut} -3 \big) \big\rfloor} g_m, %\nonumber
\label{eq. complete eigenstates}
\end{equation}
where $g_m$ are the corresponding sets of eigenstates with an odd number of cubic bricks, and $d'_m$ their multiplicities.
The spectrum, i.e. the set of all values of the $E$ parameter, $\big\{ E \big\}$, for which a nonzero eigenstate  exists, can be written as
\begin{equation}
\big\{ E \big\} = \bigcup_{m=0}^{\big\lfloor \frac{1}{6} N_{cut} \big\rfloor} \Big\{ L_{d_m}^{6m+3}(2E) = 0\Big\} \ \cup \ \bigcup_{m=0}^{\big\lfloor \frac{1}{6} \big( N_{cut} -3 \big) \big\rfloor} \Big\{ L_{d'_m}^{6m+6}(2E) = 0\Big\}. %\nonumber
\label{eq. complete eigenenergies}
\end{equation}

\subsubsection{Infinite cut-off solutions}

Eventually, one can show that the solutions retain their structure in the infinite cut-off limit.
The proof relies on the observation that the mixing coefficients $\Gamma^E(m,p)$ do not
depend on $N_{cut}$. Therefore, in order to obtain
the exact solution, the sum over $j$ can be safely extended to infinity,
\begin{align}
|E\rangle &= \sum_{j=0}^{\infty} L_j^{6m+3}(2E) \Big( |j,2m\rangle + \sum_{p=1}^{m} \Gamma^E(m,p) | j+3p,2m-2p\rangle \Big), \nonumber
\end{align}
where $E$ can be now any real number. Notice
that in the limit $N_{cut} \rightarrow \infty$ the number of separate
sets of solutions $f_m$ will become also infinite.

\section{Conclusions}
\label{sec. Conclusions}

    In this paper we have described the cut Fock space approach to $D=2$, supersymmetric
    Yang-Mills quantum mechanics. We have briefly presented the numerical algorithm as well as
    numerical results, namely the spectra of the SYMQM system with the $SU(3)$ gauge group in the
    bosonic and $n_F=2$ sectors. Subsequently, we showed that the cut Fock space approach is also a convenient
    framework for analytic calculations. We have derived exact solutions for the SYMQM system with the $SU(2)$ gauge group
    and compared them with the original solutions of Halpern and Claudson. Then, we have applied the method to the $SU(3)$ SYMQM system
    and obtained the spectra and eigenstates in the bosonic sector. Hence, for a given cut-off,
    one can explain analytically all features of figure \ref{fig. spectrum1}.

    Our analytic results can be extended in several directions. First of all, it is possible to obtain recursive relations and solve them
    in all fermionic sectors of the model with $SU(3)$ gauge group. Obviously, an exact, complete solution of this model enables one to calculate
    the Witten's index \cite{korcyl3}. Second, the knowledge of exact solutions can be helpful for the investigation of systems with interactions.
    One can use the free solutions at finite cut-off as a starting point of perturbative expansion. In both these problems the generalization
    to other $SU(N)$ gauge groups can be achieved. Third, the exact form of the solutions in the bosonic as well as fermionic
    sectors enables one to study their large-$N$ limit, which is an important point in the investigations of the SYMQM systems.
    Last but not least, the method can be extended to higher dimensional systems, its application
    to $D=4$, SYMQM with $SU(2)$ gauge group is now being investigated.

\section*{Acknowledgments}

The Author would like to thank J. Wosiek for many discussions on the subject of this paper.

\end{document}